\documentstyle[12pt]{article}

\makeatletter
\def\eqnumsection{\@addtoreset{equation}{section}\def\theequation
{\arabic{section}.\arabic{equation}}}
\makeatother

\title{
\begin{flushright}
{\large Yaroslavl State University \\
        Preprint YARU-HE-98/01 \\
        hep-ph/9801402} \\[10mm]
\end{flushright}
       {\LARGE\bf $a \to e^+ e^-$ decay in a model} \\ 
       {\LARGE\bf with induced coupling to leptons} \\}

\author{{\Large\bf N.V.~Mikheev and O.S.~Ovchinnikov} \\[2mm]
        {\large\it
             Division of Theoretical Physics, Department of Physics,} \\
        {\large\it
             Yaroslavl State University, Yaroslavl 150000, Russia} \\
        {\large\it E-mail: \, mikheev@yars.free.net; 
                           \, ovoles@univ.uniyar.ac.ru} \\[4mm]
        {\Large\bf and} \\[4mm]
        {\Large\bf L.A.~Vassilevskaya} \\[2mm]
        {\large\it
             Moscow Lomonosov University, V-952, Moscow 117234, Russia} \\
        {\large\it E-mail: \, vasilevs@vitep5.itep.ru}}

\date{}

\begin{document}

\maketitle

\begin{abstract}
The process of electron-positron pair production by axion
propagating in an external magnetic field is investigated in a model
with only induced axion coupling to leptons. Contributions of 
the lowest and higher Landau levels are analyzed. The results we
have obtained demonstrate a strong catalyzing influence of the field.
\end{abstract}

\vspace*{40mm}

\centerline{\large will be published in {\it Mod. Phys. Lett. A}}

\thispagestyle{empty}

\newpage
	
The pseudo-Goldstone boson associated with Peccei-Quinn symmetry 
$U_{PQ}(1)$~\cite{P1}, the axion~\cite{WW}, is of interest not only 
in theoretical aspects of elementary particle physics, but in some 
astrophysical and cosmological ap\-pli\-ca\-tions as 
well~\cite{Tur,Raf1,Raf2}.
It is also known that astrophysical and cosmological considerations 
have been very effective in obtaining restrictions on the axion 
mass~\cite{Raf3}
\footnote{However in paper~\cite{Rub} a possibility to solve the CP 
problem of QCD with a heavy axion $M_a \le 1$ TeV is considered.}:

\begin{equation}
10^{-5}~{\rm eV} \leq m_a \leq 10^{-2}~{\rm eV}.
\label{eq:MA}
\end{equation}

In some astrophysical considerations it is important to take 
into account both dense star matter and intensive electromagnetic 
fields. Of particular interest become 
investigations in strong external magnetic fields when the field
strength can be of order or even more the critical value
$B > B_e$, $B_e = m^{2}_{e}/e \simeq 0.44 \cdot 10^{14}$~G.
A possible existence of astrophysical objects with such extremely 
strong magnetic fields as $B \sim 10^{15} \div 10^{17}$~G
was pointed out, for example, in~\cite{magnetar,tor}.
On the other hand, in such strong magnetic fields 
otherwise negligible processes are not only opened kinematically 
but become substantial ones as well (for example, the photon splitting 
into electron-positron pair $\gamma \to e^+ e^-$~\cite{Klep},
Cherenkov process $\nu \to \nu \gamma$~\cite{Raf4,L1}).

In recent paper~\cite{L2} we have studied the field-induced axion decay 
$a~\to e^+ e^-$ in DFSZ model~\cite{DFSZ}, where axions couple to leptons 
at tree level. It was shown that the axion lifetime could be reduced 
to $10^5 $~s in case of the decaying axion energy $\sim 1$~MeV and 
the magnetic field strength $\sim 10^{15}$~G.
In this paper we investigate the field-induced axion decay into 
electron-positron pair via a photon intermediate state 
$a \to \gamma \to e^+ e^- $ in KSVZ model~\cite{KSVZ} in which axions 
have not direct coupling to leptons. 
The reason for which this channel is opened in the 
magnetic field is that $e^+e^-$-pair can have in the field both 
time--like and space--like total momentum as it occurs in photon 
splitting $\gamma \to e^+ e^-$~\cite{Klep}.
A diagram describing $a \to \gamma \to e^+ e^-$ is shown in Fig.~1
where double lines imply the influence of the 
magnetic field in the electron wave functions. The Lagrangian 
describing the effective axion-photon coupling is:

\begin{equation}
{\cal L}_{a \gamma} =  g_{a \gamma} \; \partial_{\mu} \, A_\nu \;
\tilde F_{\nu\mu} \; a,
\label{eq:Lag}
\end{equation}

\noindent
where $g_{a \gamma}$ is a constant with the dimension $(energy)^{-1}$; 
$A_\mu$ is the 4-potential of the quantized electromagnetic field,
$\tilde F$ is the dual external field tensor. In the second order of
the perturbation theory a matrix element can be presented in the form:

\begin{eqnarray}
S & = & i e \; g_{a \gamma} \int d^4 x \, d^4 y 
\left ( {\tilde F}_{\alpha \mu}^{ext} \, \partial_{\mu} \, 
a(y) \right ) \, 
\left ( {\bar \psi(x)} \, \gamma_{\nu} \, \psi(x) \right ) 
G_{\alpha \nu} (x - y)
\nonumber \\
& = & - \frac{ e \; g_{a \gamma}}{\sqrt{2 E_a V}} \,
 \int d^4 x \, \left ( {\bar \psi}(x) \, \hat h \, \psi(x) \right )
\, e^{-i q x},
\label{eq:S1} \\
h_{\alpha} & = & (q \tilde F G (q))_\alpha =
q_\mu \tilde F_{\mu\nu} G (q)_{\nu\alpha},
\nonumber
\end{eqnarray}

\noindent where $e > 0$ is the elementary charge; $\psi(x)$ is the known
solution of the Dirac equation in a magnetic field; $q_\alpha$,
$E_a$ are the 4-momentum and the energy of the decaying axion, 
respectively. The expression for the photon propagator $G_{\alpha \beta}$ 
can be presented in a diagonal form~\cite{Shab}:

\begin{eqnarray}
G_{\alpha \beta} =  \sum_{\lambda=1}^3 \;
\frac {b_\alpha^{(\lambda)} b_\beta^{(\lambda)}}{(b^{(\lambda)})^2} \; 
\frac{ 1 }{ q^2 - \ae^{(\lambda)} }, \;\;\;\;
b_{\alpha}^{(\lambda)} b_{\alpha}^{(\lambda')} =
\delta_{\lambda \lambda'} \;
(b_{\alpha}^{(\lambda)})^2
\label{eq:G}
\end{eqnarray}

\noindent in a basis:

\begin{eqnarray}
b_{\alpha}^{(1)} & = & ( q F )_{\alpha},
\label{eq:B} \\
b_{\alpha}^{(2)} & = & ( q {\tilde F} )_{\alpha},
\nonumber \\
b_{\alpha}^{(3)} & = & q^2 (q F F )_{\alpha} - 
q_{\alpha} \cdot (q F F q),
\nonumber \\
b_{\alpha}^{(4)} & = &  q_{\alpha}.
\nonumber 
\end{eqnarray}

\noindent We note that the basis vectors 
$b_{\alpha}^{(\lambda)}$~(\ref{eq:B}) are the eigenvectors of the 
photon polarization tensor with the eigenvalues $\ae^{(\lambda)}$.
As it is seen from~(\ref{eq:S1}) only the basis vector
$b_{\alpha}^{(2)} = (q \tilde F)_\alpha$ gives a contribution to the 
decay, as $(F \tilde F)_{\alpha\beta} = 0$.

One can obtain the decay probability carrying out a non-trivial 
integration over the phase space of the electron-positron pair taking 
the specific kinematics of charged particles in the magnetic field 
into account. However the probability of $a \to e^+ e^-$ decay can be 
obtained from the imaginary part of $a \to \gamma \to a$ transition 
amplitude via the virtual photon:

\begin{eqnarray}
E_a \; W_{a \to e^+ e^-} = Im \; M_{a \to a} = 
 - g_{a \gamma}^2 \; (q {\tilde F} (Im G) {\tilde F} q).
\label{eq:W0}
\end{eqnarray}

\noindent The result of our calculations is:

\begin{eqnarray}
W = - \frac{ g_{a \gamma}^2 }{ 4 \pi \alpha E_a}
\cdot \frac{ e^2 (q {\tilde F} {\tilde F} q) \cdot  
Im \; \ae^{(2)} } { \left [ \bigg (m_a^2 -  Re \; \ae^{(2)} 
\bigg )^2 + \bigg ( Im \; \ae^{(2)} \bigg )^2 \right ] } \; ,
\label{eq:W1} 
\end{eqnarray}

\noindent where $\ae^{(2)}$ has a form of a double integral:

\begin{eqnarray}
\ae^{(2)} & = & - \frac{e^2}{ 4 {\pi}^2 } \int \limits_0^1 du 
\int \limits_0^{\infty}
\frac{ dt }{t} \bigg \lbrace \frac{ \beta t }{ \sin \beta t } \;
\bigg [ \;
 \frac{ q^2_\parallel }{2} \cos \beta t (1 - u^2)
\label{eq:Ae1} \\ 
& - & \frac{ q^2_{\perp} }{2} \left (\cos \beta ut - 
\frac{ u \sin \beta ut \; \cos \beta t }{ \sin \beta t }  \right ) \; 
\bigg ]
\; e^{ -i \Phi } - \frac{ q^2 }{2} (1 - u^2) \; e^{ -i \Phi_0 } 
\bigg \rbrace,
\nonumber \\
\Phi & = & t \left ( m_e^2 - q^2_\parallel \frac{ 1 - u^2 }{4}
\right ) + \frac{ q^2_{\perp} }{ 2 \beta } \; \frac{ \cos \beta ut -
\cos \beta t }{ \sin \beta t },
\nonumber\\
\Phi_0 & = & \Phi (B = 0) = t \left ( m_e^2 - \frac{ q^2 }{4}
(1 - u^2) \right ),
\nonumber\\
q_{\parallel}^2 & = & \frac{(q \tilde F \tilde F q)}{B^2}, \;\;\;\;
q_{\perp}^2 = \frac{(q F F q)}{B^2},
\nonumber\\
\beta & = & e B = \sqrt {e^2 (FF)/2} \; .
\nonumber
\end{eqnarray}

\vspace{5mm}

The expression~(\ref{eq:W1}) is significantly simplified in two 
limiting cases which are of interest in some astrophysical 
applications:

\begin{enumerate}
\item 
The field invariant $ \vert e^2 (F F) \vert^{1/2}$ is relatively
small parameter ($E_a^2 \gg  e B$), so electron and positron are
born in states corresponding to the highest Landau levels;

\item 
In the strong field limit $e B  \gg E_a^2 $ the field
invariant $ \vert e^2 (F F) \vert^{1/2}$ appears to be the largest
physical parameter, so the electron and the positron are born 
only on the lowest Landau level. 

\end{enumerate}

In the case $E_a^2 \gg e B$ the eigenvalue $\ae^{(2)}$~(\ref{eq:Ae1})
may be described by the expression in the crossed field limit
($\vec E \perp \vec B, E = B$)~\cite{Rit}:

\begin{eqnarray}
\ae^{(2)}
\simeq \frac{ 9 \cdot 3^{1/6} \, \Gamma^4( {2 \over 3} ) }{ 14 \pi^2 } \;
\alpha \; (1 - i \sqrt{3}) \, (e^2 qFFq)^{1/3}.
\label{eq:Ae2} 
\end{eqnarray}

\noindent Using~(\ref{eq:Ae2}) we obtain the following expression for
the decay probability:

\begin{eqnarray}
W & \simeq & \frac{ 7 \cdot 3^{1/3} \; \pi }
{ 8 \cdot 9 \cdot \Gamma^4(\frac{2}{3}) \; \alpha^2 } \cdot
\frac{ g^2_{a \gamma} }{ E_a } \cdot (e^2 qFFq)^{2/3}
\label{eq:W2} \\
& = & 2,46 \cdot 10^3 \; g^2_{a \gamma} \; (eB)^{4/3} \; 
E_a^{1/3} \; \sin^{4/3} \theta
\nonumber
\end{eqnarray}

\noindent ($\theta$ is the angle between the vectors of the
magnetic field strength ${\vec B}$ and the momentum of the axion 
${\vec q}$) and the axion lifetime:

\begin{eqnarray}
\tau^{KSVZ} \simeq 1,16 \; 
\left ( \frac{ 10^{-10} }{ g_{a \gamma} GeV } \right )^2 \;
 \left ( \frac{ E_a }{ 10 \,MeV } \right )^{-1/3} \;
\left ( \frac{ 10^{15} G }
 { B \sin{\theta} } \right )^{4/3} \; s.
\label{eq:T1}
\end{eqnarray}

\noindent For comparison we present here the lifetime for DFSZ 
axions~\cite{L2} (Fig.~2) in analogous case of the decaying axion 
energy $\sim 10$~MeV in the field of strength $\sim 10^{15}$~G:

\begin{eqnarray}
\tau^{DFSZ} \simeq 1,03 \cdot 10^6  \; 
\left ( \frac{ 10^{-13} }{ g_{ae} } \right )^2
\left ( \frac{ E_a }{ 10 \,MeV } \right )^{1/3}\;
\left ( \frac{ 10^{15} G }
{ B \sin{\theta} } \right )^{2/3} \; s,
\label{eq:T2}
\end{eqnarray}

\noindent where $g_{ae}$ is a dimensionless Yukawa coupling constant.

In another limiting case, $e B  \gg E_a^2 $, when the field
strength $B$ appears to be the largest physical parameter,
the eigenvalue $\ae^{(2)}$~(\ref{eq:Ae1}) has a form:

\begin{eqnarray}
\ae^{(2)}
\simeq \frac{2 \alpha }{ \pi } \; e B \;
\left (1 - i \;\frac{2 \pi \;m_e^2}{E_a^2 \sin^2\theta} \right ).
\label{eq:Ae3} 
\end{eqnarray}

\noindent With $\ae^{(2)}$~(\ref{eq:Ae3}) the decay probability
can be significantly simplified:

\begin{eqnarray}
W  \simeq  \frac {\pi g^2_{a \gamma}}{4 \alpha^2} \;
\frac{ e B m^2_e}{ E_a}, 
\label{eq:W3}
\end{eqnarray}

\noindent and the axion lifetime is reduced to seconds 
as in previous case:

\begin{eqnarray}
\tau^{KSVZ} \simeq 0.29 \; 
\left ( \frac{ 10^{-10} }{ g_{a \gamma} \,GeV } \right )^2 \;
 \left ( \frac{ E_a }{ 1 \, MeV } \right )\; 
\left ( \frac{ 10^{16} \, G } { B } \right ) \; s.
\label{eq:T3}
\end{eqnarray}

\noindent Let us also give here the results in
DFSZ model~\cite{L2} for the same axion and field parameters:

\begin{eqnarray}
\tau^{DFSZ} \simeq 1,4 \cdot 10^4  \; 
\left ( \frac{ 10^{-13} }{ g_{ae} } \right )^2
\left ( \frac{ E_a }{ 1 \,MeV } \right ) \;
\left ( \frac{ 10^{16} \, G } { B } \right ) \; s.
\label{eq:T4}
\end{eqnarray}

The expressions for the axion lifetime both for KSVZ hadronic axions 
Eqs.~(\ref{eq:T1}),~(\ref{eq:T3}) and DFSZ axions 
Eqs.~(\ref{eq:T2}),~(\ref{eq:T4}) demonstrate the strong catalyzing 
influence of the external field on $a \to e^+ e^-$ decay forbidden in
vacuum by the momentum conservation. 

Notice that the hadronic axions lifetime in the external field is reduced 
to seconds for the appropriate axion and field parameters 
(see Eqs.~(\ref{eq:T1}) and~(\ref{eq:T3})) while the relativistic 
axion lifetime in vacuum~\cite{Raf1} is gigantic one:

\begin{equation}
\tau^{(0)} (a \to 2\gamma) \sim 10^{44} \; 
\left ( \frac{ 10^{-10} }{ g_{a \gamma} \; GeV } \right )^2 \;
\left ( {10^{-3} \; eV \over m_a } \right )^4 \; 
\left ( {E_a \over 10 \, MeV } \right ) s.
\label{eq:T0}
\end{equation}

The results we have presented here are of interest in such astrophysical 
objects where from both components of the active medium, a 
magnetic field and plasma, the magnetic component dominates. Such 
conditions can be realized, for example, in a supernova explosion or in 
a coalescence of neutron stars when a region of order of hundred 
kilometers outside the neutrinosphere
with strong magnetic fields $\sim 10^{14} \div 10^{16}$~G~\cite{tor}
and rather rarified plasma can exist.

\vspace{5mm}

The work was supported by Grant INTAS 96-0659.

\newpage

\begin{figure}[htb]


\def\photonatomright{\begin{picture}(3,1.5)(0,0)
                                \put(0,-0.75){\tencircw \symbol{2}}
                                \put(1.5,-0.75){\tencircw \symbol{1}}
                                \put(1.5,0.75){\tencircw \symbol{3}}
                                \put(3,0.75){\tencircw \symbol{0}}
                      \end{picture}
                     }
\def\photonatomup{\begin{picture}(1.5,3)(0,0)
                             \put(-0.75,3){\tencircw \symbol{3}}
                             \put(-0.75,1.5){\tencircw \symbol{2}}
                             \put(0.75,1.5){\tencircw \symbol{0}}
                             \put(0.75,0){\tencircw \symbol{1}}
                   \end{picture}
                  }
\def\photonrighthalf{\begin{picture}(30,1.5)(0,0)
                     \multiput(0,0)(3,0){5}{\photonatomright}
                  \end{picture}
                 }
\def\photonuphalf{\begin{picture}(1.5,15)(0,0)
                      \multiput(0,0)(0,3){5}{\photonatomup}
                   \end{picture}
                  }


\unitlength 1mm
\linethickness{0.4pt}
\begin{picture}(45.00,50.00)(-55.00,00.00)
\put(20.00,25.00){\circle*{3.00}}
\put(5.00,25.00){\circle*{3.00}}
\put(20.00,26.50){\line(+1,+1){15.00}}
\put(21.50,25.00){\line(+1,+1){15.00}}
\put(21.50,25.00){\line(+1,-1){15.00}}
\put(20.00,23.50){\line(+1,-1){15.00}}
\put(28.00,33.00){\line(-1,0){4.50}}
\put(28.00,33.00){\line(0,-1){4.50}}
\put(26.00,19.00){\line(+1,0){4.50}}
\put(26.00,19.00){\line(0,-1){4.50}}
\put(5.00,25.00){\photonrighthalf}
\multiput(-20.00,25.00)(7.00,0.00){4}{\line(1,0){3.00}}
\put(5.00,25.00){\line(0,-1){4.00}}
\put(5.00,20.00){\line(0,-1){4.00}}
\put(5.00,15.00){\line(0,-1){4.00}}
\put(3.00,9.00){\line(1,1){4.00}}
\put(7.00,9.00){\line(-1,1){4.00}}
\put(-5.00,30.00){\makebox(0,0)[cb]{\large $a(q)$}}
\put(37.00,32.00){\makebox(0,0)[lc]{\large $e(p_1)$}}
\put(37.00,18.00){\makebox(0,0)[lc]{\large $e(-p_2)$}}
\put(5.00,30.00){\makebox(0,0)[cc]{\large $x$}}
\put(20.00,30.00){\makebox(0,0)[cc]{\large $y$}}
\put(5.00,5.00){\makebox(0,0)[cc]{\large $F^{ext}$}}
\end{picture}
\end{figure}

\vspace{2cm}

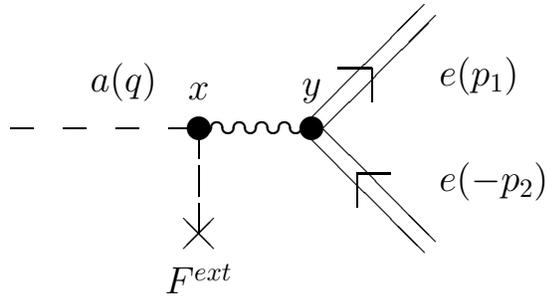
\centerline{Figure~1}

\vspace{1cm}


\begin{figure}[htb]

\unitlength 1mm
\linethickness{0.4pt}
\begin{picture}(45.00,50.00)(-45.00,00.00)
\put(20.00,25.00){\circle*{3.00}}
%
%
\put(20.00,26.50){\line(+1,+1){15.00}}
\put(21.50,25.00){\line(+1,+1){15.00}}
\put(21.50,25.00){\line(+1,-1){15.00}}
\put(20.00,23.50){\line(+1,-1){15.00}}
\put(28.00,33.00){\line(-1,0){4.50}}
\put(28.00,33.00){\line(0,-1){4.50}}
\put(26.00,19.00){\line(+1,0){4.50}}
\put(26.00,19.00){\line(0,-1){4.50}}
%
%
\multiput(-5.00,25.00)(7.00,0.00){4}{\line(1,0){3.00}}
\put(10.00,30.00){\makebox(0,0)[cb]{\large $a(q)$}}
\put(37.00,32.00){\makebox(0,0)[lc]{\large $e(p_1)$}}
\put(37.00,18.00){\makebox(0,0)[lc]{\large $e(-p_2)$}}
\end{picture}
\end{figure}

\vspace{2cm}

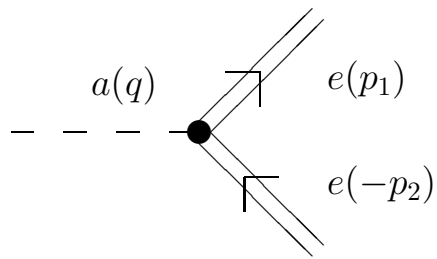
\centerline{Figure~2}

\end{document}